\documentclass[conference,a4paper]{IEEEtran}
\IEEEoverridecommandlockouts
\usepackage{amsmath}
\usepackage{cite}
\usepackage{amssymb,amsfonts}
\usepackage{algorithm}
\usepackage{algpseudocode}
\usepackage{graphicx}
\usepackage{textcomp}
\usepackage{xcolor}
\usepackage{subfigure}
\usepackage{graphicx}
\usepackage{float}
\usepackage{hyperref}
\usepackage[T1]{fontenc}
\usepackage{color}
\usepackage{setspace}
\IEEEoverridecommandlockouts

\newcommand{\alignedState}[2][.8\linewidth]{\State \parbox[t]{#1}{#2\strut}}

\newsavebox\CBox
\def\textBF#1{\sbox\CBox{#1}\resizebox{\wd\CBox}{\ht\CBox}{\textbf{#1}}}
\newcommand{\etal}{\emph{et al. }}

\def\BibTeX{{\rm B\kern-.05em{\sc i\kern-.025em b}\kern-.08em
    T\kern-.1667em\lower.7ex\hbox{E}\kern-.125emX}}
\begin{document}

\title{Unified Framework for Histopathology Image Augmentation and Classification via Generative Models}



\author{\IEEEauthorblockN{Meng Li\textsuperscript{*}, Chaoyi Li\textsuperscript{*}\thanks{* denotes equal contribution}, Can Peng, Brian C. Lovell}
\IEEEauthorblockA{The University of Queensland, School of EECS, QLD 4072, Australia}
\IEEEauthorblockA{{meng.li6}@uqconnect.edu.au, {chaoyi.li}@uq.net.au, {can.peng}@uqconnect.edu.au, {lovell}@eecs.uq.edu.au}
}

\IEEEpubid{\parbox{\columnwidth}{
    \vspace{8em} 
    \copyright 2024 IEEE. Published in the Digital Image Computing: Techniques and Applications, 2024 (DICTA 2024), 27-29 November 2024 in Perth, Western Australia, Australia. Personal use of this material is permitted. However, permission to reprint/republish this material for advertising or promotional purposes or for creating new collective works for resale or redistribution to servers or lists, or to reuse any copyrighted component of this work in other works, must be obtained from the IEEE. Contact: Manager, Copyrights and Permissions / IEEE Service Center / 445 Hoes Lane / P.O. Box 1331 / Piscataway, NJ 08855-1331, USA. Telephone: + Intl. 908-562-3966.
}\hspace{\columnsep}\makebox[\columnwidth]{ }}

\maketitle

\begin{abstract}

Deep learning techniques have become widely utilized in histopathology image classification due to their superior performance. 
However, this success heavily relies on the availability of substantial labeled data, which necessitates extensive and costly manual annotation by domain experts.
To address this challenge, researchers have recently employed generative models to synthesize data for augmentation, thereby enhancing classification model performance. 
Traditionally, this involves generating synthetic data first and then training the classification model with both synthetic and real data, which creates a two-stage, time-consuming workflow. 
To overcome this limitation, we propose an innovative unified framework that integrates the data generation and model training stages into a unified process. 
Our approach utilizes a pure Vision Transformer (ViT)-based conditional Generative Adversarial Network (cGAN) model to simultaneously handle both image synthesis and classification. 
An additional classification head is incorporated into the cGAN model to enable simultaneous classification of histopathology images.
To improve training stability and enhance the quality of generated data, we introduce a conditional class projection technique that helps maintain class separation during the generation process.
We also employ a dynamic multi-loss weighting mechanism to effectively balance the losses of the classification tasks.  
Furthermore, our selective augmentation mechanism actively selects the most suitable generated images for data augmentation to further improve performance. 
Extensive experiments on histopathology datasets show that our unified synthetic augmentation framework consistently enhances the performance of histopathology image classification models.
\end{abstract}

 \hspace*{\fill} 
 

\section{Introduction}
\label{sec:intro}

The benefits of deep learning are greatly enhanced by the availability of labeled data, which has facilitated the successful application of deep learning in histopathology image classification tasks. 
However, most of these tasks require the expertise of domain specialists and the acquisition of substantial amounts of labeled data \cite{nie2017medical}.
This process is not only labor-intensive and time-consuming but also impractical in the context of rare diseases, early-stage clinical studies, or emerging imaging modalities. 
To address these challenges, several studies have utilized generative adversarial networks (GANs) to synthesize images for data augmentation, achieving satisfactory results \cite{nie2017medical,dar2019image}. 
Despite their success, these approaches generally take to a traditional two-stage workflow, where image generation and data augmentation are treated as separate tasks. 
This separation requires distinct models for each stage, resulting in increased training complexity and effort.

Moreover, a distinct characteristic of histopathological images is the prevalence of non-local or long-range information within their composition \cite{srinidhi2021deep}. 
CNN-based generative models may struggle to synthesize realistic histopathology images since their locally focused receptive fields are not effective at capturing the non-local information in these images. 
In contrast, vision transformers (ViTs) offer promising advantages in modeling non-local context dependencies \cite{lee2021vitgan}. 
Consequently, ViT-based generative models hold significant potential for histopathology image analysis tasks.

To leverage the advantages of ViTs and transition from the two-stage paradigm to a unified approach, we propose a framework that employs a pure ViT-based GAN model for synthetic augmentation in histopathology image analysis. 
This framework integrates models from both stages into one, conditionally generating synthetic images while also performing classification prediction. 
This approach faces a significant challenge, as GANs are notorious for their training instability \cite{jiang2021transgan} and introducing multiple loss functions can further complicate this issue. 
To address these challenges, we first propose a conditioned class projection technique, designed to aid in separating conditional information during training. 
Drawing inspiration from the success of multi-task learning methods \cite{kendall2018multi,ruder2017overview}, we also introduce a multi-loss weighting function to dynamically balance the losses when training the GAN model.
Additionally, to further enhance the effectiveness of synthetic augmentation, we incorporate a selective augmentation mechanism that actively chooses the most suitable generated images for data augmentation.

The main contributions of this paper can be summarized as follows:
\begin{itemize}
\item We propose a novel unified framework that integrates the two stages of image generation and classification into a single stage, achieving higher efficiency compared to the traditional paradigm of using synthetic data for augmentation;
\item We explore the application of a conditional, pure ViT-based GAN for histopathology image analysis, highlighting its substantial practical benefits;
\item We introduce a conditioned class projection technique to improve class separation during training;
\item We propose a multi-loss weighting function that effectively stabilizes training and improves performance across tasks;
\item Extensive experiments on lymph node histopathology datasets demonstrate that our approach significantly improves classification performance.
\end{itemize}

\begin{figure*}[t]
\centering
\includegraphics[height=9cm,keepaspectratio]{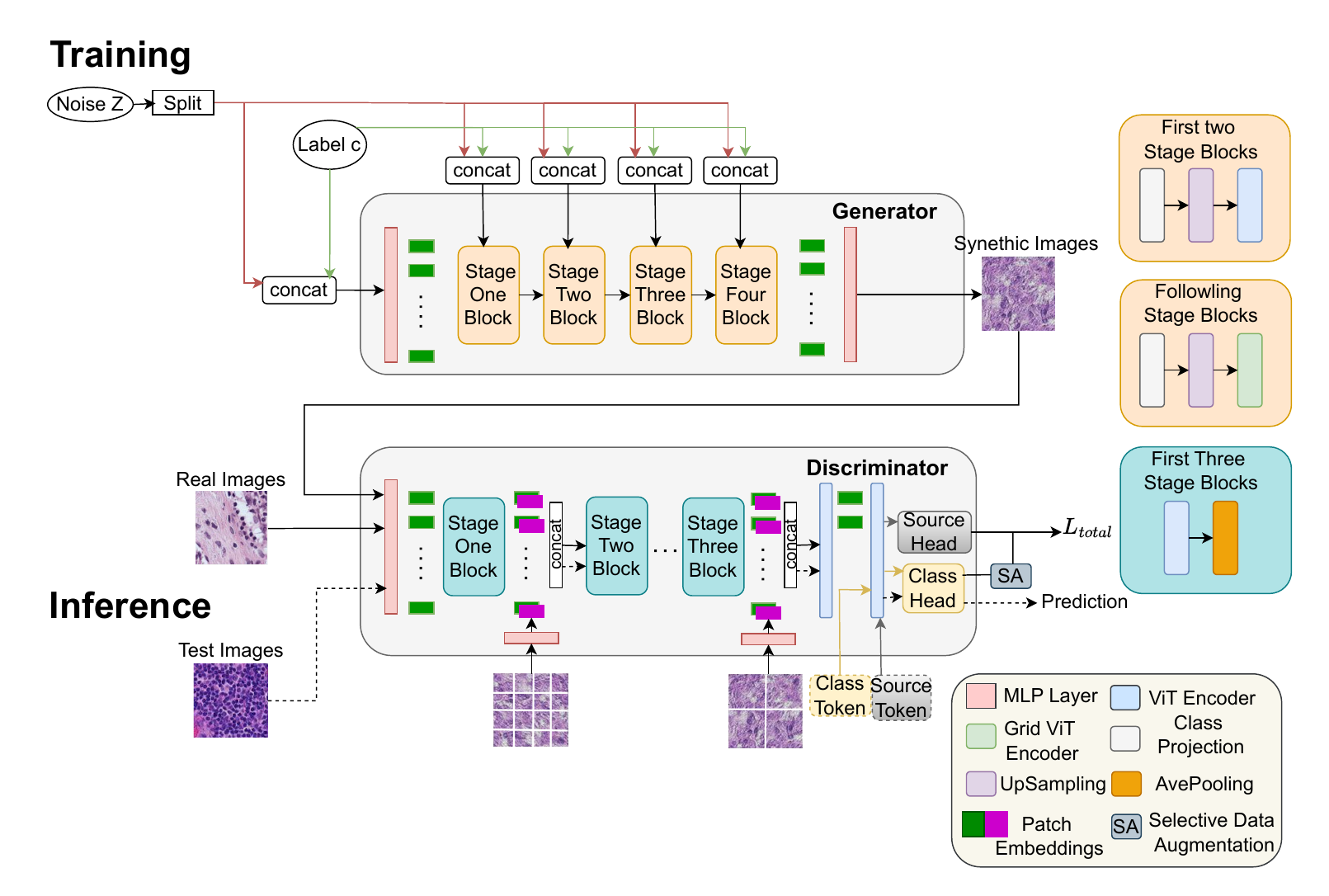}
\caption{Overview of our proposed framework. During training, In the generator, the latent vector $z$ is split into multiple chunks and concatenated with conditional label $c$ as the input to the class projection layer. A multi-scale pyramid technique is employed to learn global and local information. The selective data augmentation mechanism filters the predictions of synthetic data from the class head. Finally, the output of the discriminator is the objective function. In the inference, test data is only fed into the discriminator to obtain predictions.}

\label{fig:net_arch}
\end{figure*}
\section{Related Work}
\subsection{Generative Models for Medical Image Augmentation}
Data scarcity remains a significant obstacle in applying deep learning technologies in the medical field. 
To mitigate this challenge, data augmentation has emerged as a powerful technique for expanding training datasets. 
However, conventional augmentation methods used in general image processing often struggle or cannot be applied due to the complexities and unique characteristics of medical images.
This limitation underscores the appeal of generative models, which have demonstrated strong performance in augmenting medical data.

Currently, three primary generative models are utilized for medical image synthesis: variational autoencoders (VAEs), generative adversarial networks (GANs), and diffusion models \cite{kebaili2023deep}. 
GANs, in particular, are widely favored for their capability to generate realistic images even with limited data availability \cite{chen2022generative}, despite challenges such as training instability, convergence issues, and mode collapse \cite{mescheder2018training}.
In contrast, VAEs offer greater output diversity and avoid mode collapse compared to GANs, but they often produce blurry images, which limits their adoption for augmentation purposes. 
Recently, diffusion models \cite{sohl2015deep,ho2020denoising} have shown promise in generating realistic and diverse outputs, though they require extensive data and computational resources for training \cite{kebaili2023deep}.
Given the inherent data scarcity in medical datasets, our study focuses on augmenting data with limited availability using generative models. 
Considering these constraints, GANs emerge as the preferred choice due to their proven effectiveness in generating high-quality medical images under restricted data conditions.

\subsection{Transformer-based GANs}
A GAN model consists of two key components: a discriminator and a generator. 
During training, these components are alternatively trained. 
The discriminator learns to differentiate between the synthetic images generated by the generator and real training data.
Conversely, the generator learns to create synthetic images that are as realistic as possible to deceive the discriminator.
During testing, only the well-trained generator is used to synthesize realistic images. 

GAN was originally proposed using the convolutional neural network (CNN) architecture to formulate a min-max optimization problem aimed at narrowing the gap between real and synthetic data distributions \cite{arjovsky2017wasserstein}. 
Recently, transformer networks \cite{devlin2018bert} have attracted attention for their ability to effectively model non-local contextual dependencies.
In the realm of medical image synthesis, there is a growing trend towards replacing CNNs with transformer networks, given the significant role of global understanding of training data in generating realistic medical images. 
For instance, Korkmaz \etal \cite{korkmaz2021deep} introduced the GVTrans model, which employs cross-attentive visual transformers to map low-dimensional noise and latent variables onto MRI data. 
Zhang \etal \cite{zhang2022ptnet3d} proposed PTNet3D, leveraging pyramid transformer networks to generate high-resolution 3D longitudinal infant brain MRI data.
However, these methods often adopt a hybrid structure that integrates a transformer-based generator alongside a CNN-based discriminator.

In contrast to hybrid models, a pure transformer-based GAN offers architectural simplicity that can potentially reduce method complexity, thereby facilitating stable training and reducing computational overhead.
MedViTGAN stands out as a pure vision transformer-based conditional GAN designed specifically for generating histopathology images for data augmentation \cite{li2022medvitgan}.
Inspired by MedViTGAN, our study adopts a pure transformer GAN framework to capture non-local or long-range information in histopathology images.
Furthermore, different from previous methods, our framework integrates conditional class projection and selective data augmentation techniques to enhance both class separation and generation quality.

\subsection{Enhancing GAN Performance with Multi-Loss Integration}
To enhance performance in various image synthesis applications, several GAN-based studies have explored combining the Wasserstein loss with different loss functions. 
For instance, Liu \etal \cite{liu2018auto} integrated total variation loss, pixel loss, and feature-level losses linearly with the Wasserstein loss function to train an auto-painter, resulting in improved painting quality.
Similarly, Ebenezer \etal \cite{ebenezer2019single} employed a combination of Wasserstein loss, L1 pixel loss, and VGG loss to train their GAN.
However, these approaches typically involve manual tuning of the weights assigned to each loss component.
This manual hyperparameter tuning relies heavily on expert knowledge and can lead to time-consuming and computationally expensive training processes. 
Moreover, it may limit the generalizability of the method and result in sub-optimal performance.

The success of multi-task learning suggests a more sophisticated approach to combining loss functions across different tasks. 
Kendall \etal \cite{kendall2018multi} introduced a multi-task loss formulation that leverages the homoscedastic uncertainty of each task to dynamically weigh their respective losses. 
Inspired by their work, we propose a multi-loss weighting function that automates the adjustment of these weights during the training process.
\section{Method}
\label{sec:method}
The overall framework of our method, illustrated in Fig.~\ref{fig:net_arch}, aims to integrate image generation and classification into a unified stage. 
We build upon the transformer-based GAN architecture introduced in MedViTGAN \cite{li2022medvitgan} as the baseline model in our approach. 
This transformer structure effectively captures essential global contextual dependencies and simplifies training.

During the training stage, we introduce a conditional training strategy to enhance the model's capabilities. 
This strategy incorporates multiple class projection modules within the generator and adds an extra classification head to the discriminator.
The discriminator is equipped with two classification heads: one source head for distinguishing real and synthetic images, and one class head for image class classification. 
We then utilize a multi-loss weighting function to train the generator and the discriminator.
We further refine the synthetic augmentation process by implementing a selective data augmentation mechanism, which filters and prioritizes high-quality synthetic images for augmentation.
In the inference stage, classification tasks are solely performed by the discriminator. 
The pseudo-code outlining the training and inference procedures is provided in the Algorithm \ref{alg:trainging_inference}.

\begin{algorithm}
\caption{Training and inference of the proposed framework}
\label{alg:trainging_inference}
\begin{algorithmic}
\State \textbf{Training Phase}
\State \textbf{Input:} Truncation value $\tau$, Class Label $c$, Training data $I_{real}$, Conditional Class Projection Module $CP$, Selective Data Augmentation Mechanism $SA$, Objective Function $\mathcal{L}_{total}$, Generator $G$, Discriminator Feature Extractor $D$ with Source Head $f^{1}$ and Class Head $f^{2}$
\State \textbf{Output:} Generator $G$, Discriminator Feature Extractor $D$ with $f^{1}$ and $f^{2}$
\State Initialize $G$, $D$ with $f^{1}$ and $f^{2}$
\For{$epoch = 1$ to $N$}
    \alignedState{A noise vector $z \sim \text{TruncNormal}(0, 1, -\tau, \tau)$}
    \alignedState{$I_{\text{syn}} = G(CP(z, c))$}
    \alignedState{$y_{1}, y_{2} = f^{2}(D(I_{\text{real}}, I_{\text{syn}}))$}
    \alignedState{$y_{3} = f^{1}(D(I_{\text{real}}, I_{\text{syn}}))$}
    \alignedState{$y_{2}^{\text{filtered}} = SA(y_{2})$}
    \alignedState{Compute $\mathcal{L}_{total}$ using $y_{1}$ , $y_{2}^{\text{filtered}}$, $y_{3}$ to train $G$ and $D$}
\EndFor
\State Return $G$, $D$ with $f^{1}$ and $f^{2}$

\alignedState{ \Comment{$y_{1,2,3}$ represent classification predictions of real and synthetic data from the class head and real/synthetic predictions from the source head.}}
\\
\State \textbf{Inference Phase}
\State \textbf{Input:}  Test data $I_{test}$, Trained  Discriminator Feature Extractor $D$ with Class Head $f^{2}$ 
\State \textbf{Output:} Prediction $y_{t}$
\State $y_{t} = f^{2}(D(I_{\text{test}}))$
\State Return $y_{t}$
\end{algorithmic}
\end{algorithm}

\subsection{The Architecture of Transformer-based GAN}
The GAN architecture utilized in our framework is a pure transformer-based model inspired by TransGAN~\cite{jiang2021transgan}. 
The core component is the Vision Transformer (ViT) encoder~\cite{dosovitskiy2020image}, comprising multi-head self-attention modules and a feed-forward multi-layer perceptron (MLP) with GELU non-linearity. 
The self-attention modules and MLPs in the ViT encoder are equipped with residual connections and layer normalization for improved stability and performance.

Our GAN's generator is structured with four-stage blocks designed to progressively learn and upsample the given latent vector. Each stage block consists of a class projection layer, an upsampling module, and four ViT encoders. As the generator advances through these stages, it incrementally increases the feature map resolution until it reaches the target dimensions. Initially, the latent noise vector $z$ is concatenated with the one-hot class label $c$ and passed through a linear projection layer to generate embedding tokens $X_{0} \in \mathbb{R}^{H_{0} \times W_{0} \times C}$. These tokens are then sent to the first stage of the generator.
In the first stage, the embedded feature map undergoes upsampling from $X_{0} \in \mathbb{R}^{H_{0} \times W_{0} \times C}$ to $X_{1} \in \mathbb{R}^{2 H_{0} \times 2 W_{0} \times C}$ using cubic interpolation, ensuring early feature learning without reducing dimensions. 
In the following three stages, a pixel-shuffle module~\cite{shi2016real} is used to upsample the resolution by a factor of 2$\times$, while decreasing the channels to one-quarter. 
This approach reduces memory requirements and enhances network efficiency.  
To tackle the issue where the self-attention module sacrifices local information to learn global correspondences during the high-resolution generation stage (the third and fourth stages), we replace the self-attention module with the grid self-attention module \cite{jiang2021transgan} in the ViT encoder.
This modification enables our model to effectively capture both global and local information.
Specifically, we set the predefined window size to $32 \times 32$ in the third stage and $16 \times 16$ in the fourth stage.

The discriminator is similarly structured into four stages. 
The first three stages consist of a ViT encoder and an average pooling layer. 
The final stage includes two ViT encoders that handle the class and source heads, respectively. 
Between the first three stages, we apply a multi-scale technique that combines the outputs from the last block with patches of varying sizes extracted from the same input image.
After passing through an MLP layer, the patch information is encoded into a sequence of embeddings for concatenation, allowing the model to learn both semantic structure and texture details efficiently.

For conditional learning, a class token and a source token are appended to the beginning of the 1D sequence at the end of the third block. 
This modified sequence then passes through the fourth stage, enabling the model to make both class predictions and real/synthesis judgments.

\subsection{Conditioned Class Projection}
To further improve the conditional image generation, we propose the conditional class projection module, including the conditional direct skip connection and the class projection layer.
Inspired by the skip-z approach \cite{brock2018large}, we propose conditional skip-z, which integrates skip-z with class-specific information. 
Skip-z \cite{brock2018large} feeds the noise vector $z$ into multiple layers of the generator, instead of just the initial layer. 
This design enables the generator to utilize the diversity of input noise to enhance the quality and diversity of the generated images. 
In our approach, we divide $z$ into chunks corresponding to each stage and concatenate each chunk with the class vector $c$. 
The combined vector $[c, z]$ is then processed through the class projection layer at each stage, enhancing both the quality and diversity of the synthetic images within the class.

Next, we project $[c, z]$ onto token embeddings by the class projection layer which is a conditional layer normalization.
Unlike class-conditional methods that use batch normalization \cite{miyato2018cgans,brock2018large,kang2020contragan}, we find that layer normalization~\cite{ba2016layer} yields superior performance for ViT-based conditional GANs. 
This is possibly explained by the fact that layer normalization can preserve positional information learned by the attention mechanism, whereas batch normalization may compromise this information \cite{xiong2020layer}.
The approach transforms a layer’s activations $a$ into a class-specific normalized activation $\bar{a}$, which is described as:
\begin{equation}\label{eq:cls}
\bar{a}=\frac{a-\hat{\mu}}{\sqrt{\hat{\sigma}+\epsilon}} * \gamma+\beta.
\end{equation}

where $\hat{\mu}$ and $\hat{\sigma}$ represent the mean and variance of the input, and $\epsilon$ ensures numerical stability. 
We introduce class-conditional information through parameters $\gamma$ and $\beta$ as linear transformations of the class embedding $c$, where $\gamma := W_\gamma^\top[c, z]$ and $\beta := W_\beta^\top[c, z]$, with $[c, z]$ denoting the concatenation of $c$ and $z$ in the conditional direct skip-z.

\subsection{Multi-Loss Weighing Function}
Our framework's learning strategy is inspired by AC-GAN \cite{odena2017conditional}, which incorporates an auxiliary class head in the discriminator.
However, the auxiliary classifier in \cite{odena2017conditional} primarily functions to guide diversified image generation, often resulting in suboptimal classification performance.
To address this issue, we introduce a multi-loss weighting function that integrates the WGAN-GP loss \cite{gulrajani2017improved} with the classification cross-entropy loss.
This approach ensures that both the generation and classification aspects of the model perform satisfactorily.
We first apply the WGAN-GP loss to our GAN. The objective function is defined as follows:

\begin{equation}
\begin{aligned}
&\mathcal{L}_{S} = \min _{G} \max _{f^{1}, D} \underset{I \sim \mathbb{P}_{r}}{\mathbb{E}}[f^{1}(D(I))]-\underset{z \sim \mathbb{P}_{z}}{\mathbb{E}}[f^{1}(D(G(c,z)))]\\
 &\quad+\lambda \underset{\hat{I} \sim \mathbb{P}_{\hat{I}}}{\mathbb{E}}\left[\left(\left\|\nabla_{\hat{I}} f^{1}(D(\hat{I}))\right\|_{2}-1\right)^{2}\right], \label{wgangp-loss}
\end{aligned}
\end{equation}

where $f^{1}$ is the source head of the discriminator, $D$ denotes the discriminator feature extractor and $G$ is the generator.
$\mathbb{P}_{r}$, $\mathbb{P}_{z}$, and $\mathbb{P}_{\hat{I}}$ donate the distribution of real data $I_{real}$, the normal distribution of a random noise vector $z$, and the distribution of pairs of points from $\mathbb{P}_{r}$ and $\mathbb{P}_{z}$. 
 Each generated sample is associated with a class label $c \sim \mathbb{P}_{c}$ besides $z$. 
$G$ uses conditioned noise to generate images $I_{\text {syn}}=G(c, z)$. 
Note that the source head does not involve conditioned information. 

To enable conditional learning, we employ a class head $P(C \mid I)$ on the discriminator, the log-likelihood objective function is given by:

\begin{equation}\label{eq:dclsreal}
\begin{aligned}
\mathcal{L}_{C}=\mathbb{E}\left[\log p\left(C=c \mid I_{\text {real }}\right)\right]+ \mathbb{E}\left[\log p\left(C=c \mid I_{\text {syn}}\right)\right], 
\end{aligned}
\end{equation}

both $D$ and $G$ are trained to maximize $\mathcal{L}_{C}$. 
We find that simply combining $\mathcal{L}_{C}$ and $\mathcal{L}_{S}$ is prone to result in training failure as the WGAN-GP loss value is relatively large at most times, which results in cross-entropy loss overwhelmed during training. 
In addition, the model tends to perform well on the discriminator for a particular classification task, leading to an imbalance problem. 

To this end, we extend the concept from the multi-task learning realm~\cite{kendall2018multi} to weigh classification losses and propose a multi-loss weighing loss function, which balances losses of multi-classification tasks by considering the uncertainty of each task. 
Specifically, let $f^{2}(I)$ be the output of the class head and $\exp \left(-\sigma\right)$ be the trainable scaling parameter, our adapted classification likelihood of the model output through the softmax function can be written as:

\begin{equation}
	p\left(y \mid f^{2}(I), \sigma\right)=\operatorname{Softmax}\left(\exp(-\sigma) f^{2}(I)\right), \label{2}
\end{equation}

with $\sigma \in [-\infty, +\infty]$. 
The scaling process can be regarded as a Maxwell–Boltzmann distribution, where $\exp(-\sigma)$ is commonly referred to as the temperature of the input. 
The learnable parameter's magnitude determines how ``flat'' the discrete distribution is. 
The output is then related to uncertainty by using the log-likelihood, which can then be written as:

\begin{small}
\begin{equation}
	\begin{aligned}
         -\operatorname{log}p\left(y=c\mid f^{2}(I),\sigma\right)
		&=-\operatorname{logSoftmax}\left(\exp(-\sigma)f^{2}_{c}\left(I\right)\right)\\
		&= -\operatorname{log}\frac{\exp[\exp(-\sigma)f^{2}_{c}(I)]}{\sum_{c'}\exp[\exp(-\sigma)f^{2}_{c'}(I)]}\\
        &=-\exp(-\sigma)\operatorname{log}[\frac{\exp({f^{2}_{c}(I))}}{\sum_{c'}\exp({f^{2}_{c'}(I))}}]\\
        & \,\quad +\operatorname{log}\frac{\sum_{c'}\exp[\exp(-\sigma)f^{2}_{c'}(I)]}{\left(\sum_{c'}\exp({f^{2}_{c'}(I))}\right)^{\exp(-\sigma)}}\\   
        & \approx \exp(-\sigma)\mathcal{L} + \sigma, \label{eq3}     
	\end{aligned}
\end{equation}
\end{small}

we write $\mathcal{L}= -\operatorname{logSoftmax}\left(y,f^{2}(I)\right)$ for the cross entropy loss (not scaled). 
Then, an explicit simplifying assumption is introduced as $\exp(-\sigma) \sum_{c'} \exp \left[\exp(-\sigma) f_{c'}^{2}(I)\right] \approx\left(\sum_{c'} \exp \left(f_{c'}^{2}(I)\right)\right)^{\exp(-\sigma)}$. 
When $\sigma \to 0$, this equation becomes equality. 
This helps simplify $\operatorname{log}\frac{\sum_{c'}\exp[\exp(-\sigma)f^{2}_{c'}(I)]}{\left(\sum_{c'}\exp({f^{2}_{c'}(I))}\right)^{\exp(-\sigma)}}$ to  $\sigma$ and empirically demonstrates promising results.

\begin{singlespace}
Given multiple classification outputs from the class head, we often define the likelihood of factorizing over the outputs. Our multi-task likelihood is shown as follows:
\end{singlespace}

\begin{footnotesize}
\begin{equation}
    \begin{aligned}
    p\left(y_{1},y_{2} \mid f^{2}(I_{real,syn})\right)=
    &p\left(y_{1}\mid f^{2}(I_{real})\right)p\left(y_{2}\mid f^{2}(I_{syn})\right), \label{1}
    \end{aligned}
\end{equation}
\end{footnotesize}

where $y_{1,2}$ represent classification predictions of real and synthetic data from the class head, respectively.
Next, the joint loss of weighted log-likelihood of multi-classification tasks is given as:
\begin{small}
\begin{equation}
	\begin{aligned}
		\mathcal{L}_{mlw}
		&=-\operatorname{log}p\left(y_{1},y_{2}=c_{1},c_{2}\mid f^{2}(I_{real,fake}),\sigma_{1},\sigma_{2}\right)\\
        &=\exp(-\sigma_{1})\mathcal{L}_{1}+\exp(-\sigma_{2})\mathcal{L}_{2}+\sigma_{1}+\sigma_{2}
	\end{aligned}
\end{equation}
\end{small}

$L_{mlw}$ can be regarded as learning the weights of each output loss.
A larger temperature value, $\sigma$, results in a reduced contribution to the loss function, whereas a smaller $\sigma$ increases its weight. 
When $\sigma$ is very small, the function is predominantly regulated by the last three terms. 
Our experiments indicate that the range of $\sigma$ values impacts model performance during training. 
Consequently, differently from the approach of \cite{kendall2018multi}, we utilize $\exp \left(-\sigma\right)$ as weights for different losses. 
This allows $\sigma$ to act as an infinitely regularized value, enhancing the model's performance by providing a more balanced loss function across various tasks.
The total loss $L_{total}$ is shown as the following:

\begin{equation}
    \mathcal{L}_{total} = \mathcal{L}_{S} + \mathcal{L}_{mlw}
\end{equation}

\subsection{Selective Data Augmentation}
We further enhance the performance of synthetic augmentation by employing a selective data augmentation mechanism. 
To ensure high-fidelity image generation, we first use a truncation method that resamples $z$ from a truncated normal distribution with a truncation value $\tau$. After generating the images conditionally, we leverage the class head to select high-quality augmentation images. 
Only generated images with prediction values exceeding a threshold $\lambda$ will be used to compute the classification loss function. 
This approach helps to mitigate the negative impact of unrealistic synthetic data on model training.
Our experiments indicate that setting $\tau = 0.7$ and $\lambda = 0.6$ provides the optimal performance.
The pseudo-code of selective data augmentation is shown in the Algorithm \ref{alg:selective_augmentation}.

\begin{algorithm}
\caption{Selective Data Augmentation Mechanism}
\label{alg:selective_augmentation}
\begin{algorithmic}
\State \textbf{Input:} Truncation value $\tau$, Confidence threshold $\lambda$, Conditional Class Projection Module $CP$, Class Label $c$, Generator $G$, Discriminator Feature Extractor $D$ with Source Head $f^{1}$ and Class Head $f^{2}$
\State \textbf{Output:} Filtered prediction $y_{2}^{\text{filtered}}$
    \State A noise vector $z \sim \text{TruncNormal}(0, 1, -\tau, \tau)$
    \State $I_{\text{syn}} = G(CP(z, c))$
    \State $y_{2} = f^{2}(D(I_{\text{syn}}))$
    \If{$y_{2} > \lambda$}
        \State $y_{2} \to y_{2}^{\text{filtered}}$
    \Else
    \State $y_{2}$ is discarded
    \EndIf
\State \Return $y_{2}^{\text{filtered}}$
\end{algorithmic}
\end{algorithm}
\begin{figure*}[t]
  \centering
  \includegraphics[width=18cm,keepaspectratio]{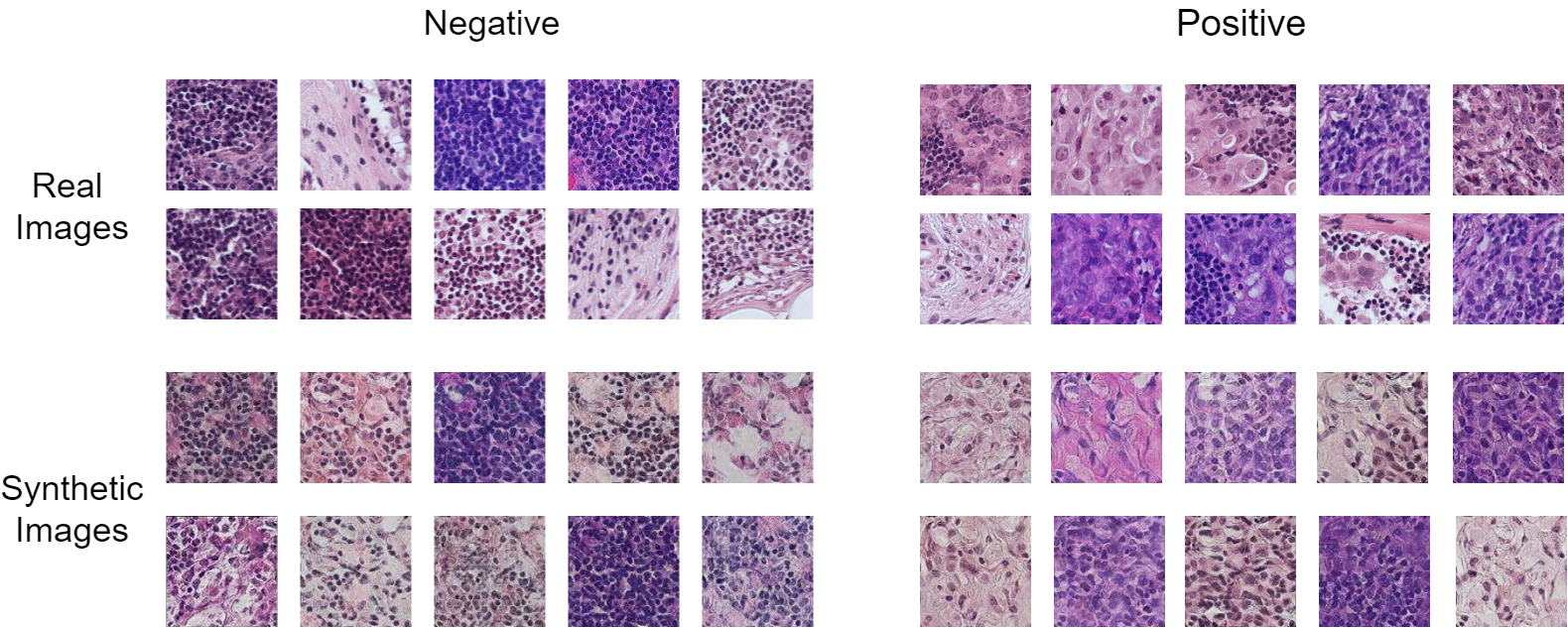}
  \caption{Visualization comparison between real images from the training data and images generated by the proposed model.
  }
  \label{fig:gen_imgs}
\end{figure*}

\begin{table*}[t]
	\centering
	\caption{Comparison of various baseline methods with and without our data synthesis augmentation method on the PCam dataset~\cite{veeling2018rotation}.
 }
	\label{table:results}
	\scalebox{0.9}{
		\begin{tabular}{l|c|c|c|c} 
			\hline
			\begin{tabular}[c]{@{}l@{}}\\\end{tabular} & Accuracy                    & AUC                         & Sensitivity                 & Specificity                  \\ 
			\hline
			ResNet34~\cite{he2016deep}                                  & 0.881 $\pm$ 0.079           & 0.945 $\pm$ 0.022           & 0.846 $\pm$ 0.086           & 0.916 $\pm$ 0.020            \\
			ResNet34 + Synthetic Data                       & 0.916~$\pm$ 0.144 & 0.954~$\pm$ 0.042           & 0.891~$\pm$ 0.100 & 0.951~$\pm$ 0.109  \\ 
			\hline
			ResNet50\_CBAM~\cite{woo2018cbam}                            & 0.899 $\pm$ 0.131           & 0.955 $\pm$ 0.045           & 0.863 $\pm$ 0.096           & 0.935 $\pm$ 0.040            \\
			ResNet50\_CBAM + Synthetic Data                  & 0.922~$\pm$ 0.077  & 0.962 $\pm$ 0.046  & 0.879~$\pm$ 0.047           & 0.945~$\pm$ 0.110   \\ 
			\hline
			DenseNet169~\cite{huang2017densely}                                & 0.894 $\pm$ 0.059           & 0.955 $\pm$ 0.036           & 0.881 $\pm$ 0.094           & 0.908 $\pm$ 0.032            \\
			DenseNet169 + Synthetic Data                     & 0.928~$\pm$ 0.076 & 0.960~$\pm$ 0.049           & 0.898~$\pm$ 0.125  & 0.967~$\pm$ 0.092   \\ 
			\hline
			MedViTGAN~\cite{li2022medvitgan}                  & 0.939 $\pm$ 0.059           & 0.980 $\pm$ 0.073           & 0.906 $\pm$ 0.079           & 0.974 $\pm$ 0.054            \\
            \hline
			Ours        & \textBF{0.945~}$\pm$\textBF{ 0.054 } & \textBF{0.981~$\pm$ 0.045 } & \textBF{0.910~$\pm$ 0.063 } & \textBF{0.977~$\pm$ 0.102 }  \\
			\hline
		\end{tabular}
	}
\end{table*}

\begin{table*}[t]
    \centering
    \caption{Ablation study of the proposed approach on the PatchCamelyon benchmark dataset. 
    CP: conditional class projection. 
    SA: selective data augmentation.}
    \label{table:1}
    \scalebox{0.9}{
    \begin{tabular}{c|c|c|c|c|c|c}
    \hline
        Baseline & CP & SA & Accuracy & AUC & Sensitivity & Specificity \\ \hline
       $\checkmark$ & ~ & ~ & 0.916 $\pm$ 0.070  & 0.971 $\pm$ 0.098 & 0.900 $\pm$ 0.033 & 0.923 $\pm$ 0.081 \\ 
        $\checkmark$ & $\checkmark$ & ~ & 0.931 $\pm$ 0.029 & 0.966 $\pm$ 0.101 & 0.897 $\pm$ 0.277 & 0.957 $\pm$ 0.123 \\ 
        $\checkmark$ & ~ & $\checkmark$ & 0.929 $\pm$ 0.110  & 0.977 $\pm$ 0.093 & 0.889~$\pm$ 0.042 & 0.961 $\pm$ 0.069 \\
        $\checkmark$ & $\checkmark$ & $\checkmark$ & \textBF{0.945~}$\pm$\textBF{ 0.054 } & \textBF{0.981~$\pm$ 0.045 } & \textBF{0.910~$\pm$ 0.063 } & \textBF{0.977~$\pm$ 0.102 } \\ \hline
    \end{tabular}
    \label{table_ab}
    }
\end{table*}

\section{Experiments}
\label{exp}
In this section, we present experiments conducted on a benchmark dataset to validate the effectiveness of the proposed method. 
Additionally, we perform an ablation study to evaluate the contribution of each component of our approach.

\subsection{Datasets}
We use a public PatchCamelyon (PCam) benchmark dataset~\cite{veeling2018rotation} for our experiments. The PCam dataset consists of 327,680 color images obtained from lymph node sections, digitized with a 40x objective, resulting in a pixel resolution of 0.243 microns. Each image patch has a resolution of 96 $\times$ 96 pixels and is classified into two categories based on the presence of metastatic tissue in the central region. The dataset is split into 75\% for training, 12.5\% for validation, and 12.5\% for testing, using a hard-negative mining regime. To simulate scenarios with limited training data, we randomly select only 10\% of the training images (32,768 images). The entire test set is used to evaluate model performance.

\subsection{Implementation Details}
All experiments are conducted using two Tesla V100 GPUs, each with 16GB of RAM, and implemented with PyTorch. 
An Adam optimizer with parameters $\beta_{1}=0$, $\beta_{2}=0.99$, and a learning rate of $1e-4$ is used to tune both the generator and discriminator. 
The batch size is set to 12 for both the generator and discriminator.  
Training is performed over 500 epochs for all experiments.
We select the best epoch based on two criteria: the Frechet Inception Distance (FID) score between the generated images and the validation data, and the classification performance on the validation set.

\subsection{Evaluation Metrics}
To comprehensively evaluate the performance of the trained classifier, we employed four evaluation metrics in our experiments: accuracy, area under the ROC curve (AUC), sensitivity, and specificity.
Accuracy is useful for datasets with a balanced distribution of categories but can be misleading when categories are unbalanced. 
AUC, on the other hand, provides an assessment of the model's overall performance across different thresholds and is suitable for datasets with unbalanced categories. Sensitivity is critical in tasks where identifying all positive samples is essential, as it measures recall. 
Specificity is important in tasks requiring precise identification of negative samples, as it measures the true negative rate.
This combined assessment offers a thorough evaluation of the model's performance in terms of its accuracy and its ability to correctly identify both positive and negative samples.

\subsection{Main Results}
We conducted a comparative analysis between our proposed method, MedViTGAN~\cite{li2022medvitgan}, and several representative baseline models under identical settings. 
The baseline models include ResNet34~\cite{he2016deep}, ResNet50 CBAM~\cite{woo2018cbam}, and DenseNet169~\cite{huang2017densely}. 
For the baseline models, we present results both with and without utilizing synthetic data generated by our proposed method during training.
Each model was executed five times with random initialization to ensure a fair comparison. 
Mean and standard deviation values of the results are reported for comprehensive evaluation.

Table.~\ref{table:results} presents the experimental results. 
Comparing the results of baseline models with and without synthetic data demonstrates a significant improvement in classification performance when using data generated by our method. 
For instance, compared to the baseline DenseNet169, the DenseNet169 trained with synthetic data shows an improvement of 3.4\%, 0.5\%, 1.7\% and 5.9\% in accuracy, AUC, sensitivity and specificity, respectively.
Furthermore, our proposed method surpasses MedViTGAN in terms of classification performance. 
As illustrated in Fig.~\ref{fig:gen_imgs}, the synthetic images exhibit comparable quality in fidelity and diversity to real training images. 
Thus, we believe that our proposed model is effective for handling challenges related to limited training data in histopathology image tasks.

\subsection{Ablation Study}

To further validate the effectiveness of our proposed method, we conducted an ablation study to assess the contributions of its two main components: the conditional class projection module and selective data augmentation. 
The findings are presented in Table \ref{table:1}.
First, we removed both the conditional class projection and selective data augmentation from the framework to establish a baseline. 
Then, we gradually added these two modules back into the framework to analyze their individual and combined effects.
Compared to the baseline, adding the conditional class projection improved accuracy by 1.5\% and specificity by 3.4\%, although it slightly underperformed the baseline in terms of AUC and sensitivity. 
Similarly, incorporating the selective data augmentation mechanism resulted in a 1.3\% improvement in accuracy, 0.6\% increase in AUC, and 3.8\% enhancement in specificity over the baseline, while maintaining sensitivity at nearly the same level as the baseline.
The decrease in AUC and sensitivity when using only conditional class projection may be due to the generation of some noisy synthetic data, slightly diminishing the model's performance in these areas.
Using only selective data augmentation avoids this issue, but its improvements in accuracy and specificity are not as pronounced as those achieved with conditional class projection alone.
Overall, the results of our proposed method showed a 3.2\% increase in accuracy over the baseline, demonstrating the effectiveness of our framework in enhancing classification performance.

\begin{figure}[t]
  \centering
  \includegraphics[height=5cm,keepaspectratio]{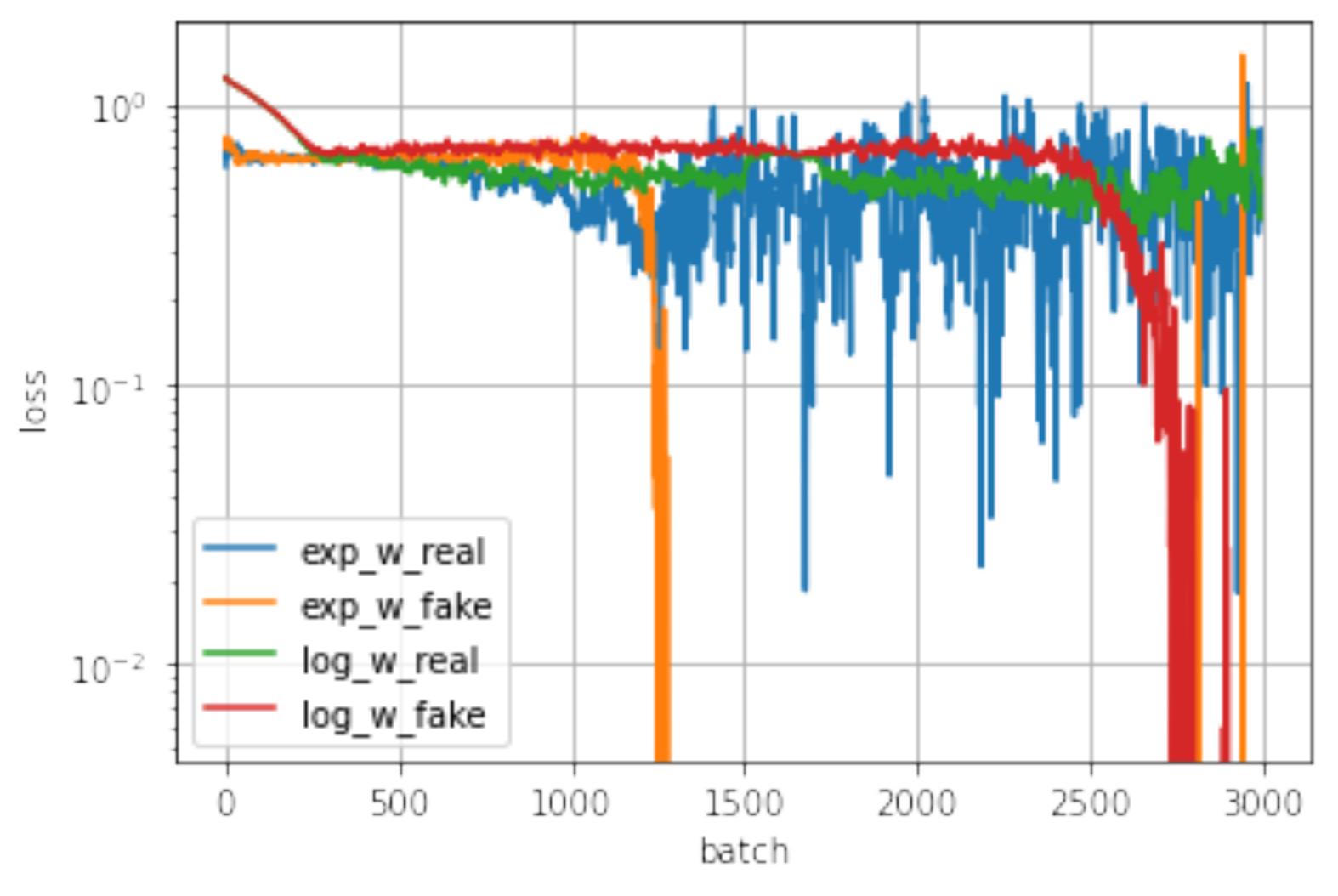}
  \caption{Comparison of training loss performance between our proposed multi-task loss function (exp\_) and the multi-task loss function by Kendall \etal \cite{kendall2018multi} (log\_).
  }
  \label{fig:1}
\end{figure}

Additionally, we analyzed the impact of different loss functions on the generated images. 
In these experiments, all settings were kept constant except for the loss function. 
We compared the multi-task loss function proposed by Kendall \etal \cite{kendall2018multi} (denoted as log\_) with our multi-loss weighing function (denoted as exp\_).
The results in Fig.\ref{fig:1} demonstrate that with our loss function applied, the fake loss converges rapidly, indicating that the learning of feature separation achieves robust results.

\section{Conclusion}

In this paper, we introduce a unified framework that integrates the two stages of image generation and classification into a single stage for the synthetic augmentation of histopathology images. 
To facilitate conditional learning, we incorporate an auxiliary discriminator head for classification.  
Additionally, we integrate a conditional class projection and a selective data augmentation mechanism to enhance class separation and improve the quality of generated images.
Furthermore, we propose a multi-loss weighing function to stabilize training and boost classification performance. 
Experimental results demonstrate that our approach significantly outperforms baseline methods. 
Future work will focus on extending our framework to other generative models, such as diffusion models.


\bibliographystyle{IEEEtran}  

\end{document}